\documentstyle[11pt]{article}
\input FEYNMAN
\title{\vspace{-1.in} \hfill {\small\rm TUM-HEP-308/98} \\
~\\~\\
Mirror Families in Electro-weak Symmetry Breaking}
\author{ Manfred Lindner
\thanks{e-mail:Manfred.Lindner$@$physik.tu-muenchen.de} 
 $\;$ and George Triantaphyllou
\thanks{e-mail:George.Triantaphyllou$@$physik.tu-muenchen.de}
$\;$\\~  \\{\it Institut f\"ur Theoretische Physik, Technische 
Universit\"at M\"unchen}\\
{\it James-Franck-Strasse, D-85748 Garching, GERMANY }}
\begin{document}
\setlength{\baselineskip}{20pt}
\maketitle
\thispagestyle{empty}  
\begin{abstract}
We study symmetry breaking in a left-right symmetric extension of
the Standard Model with mirror fermions, one for each Standard-Model
fermion. 
The new particles assist a top-quark 
condensate in breaking electro-weak symmetry. 
Half of the fermions acquire electro-weak-invariant masses at
around $500$ GeV and would be probably accessible
at future high-energy experiments like LHC or NLC. 
The contributions to
the $S$ and $T$ parameters are small and negative in accordance with
electro-weak precision data.  
\end{abstract}
\hspace{2.in} To appear in {\it Physics Letters} {\bf B} 
\vspace{2.in}
\setcounter{page}{0}
\pagebreak

\section{Introduction}
High-energy experiments have given so far data consistent with the
Standard Model
described by the gauge group structure
$SU(3)_{C} \times SU(2)_{L} \times U(1)_{Y}$. 
 However, it is well
known that the conventional
Higgs mechanism implemented for the $SU(2)_{L}$ symmetry breaking has a 
naturalness problem, in that it is hard to keep a mass of a fundamental
scalar at energies as low as the weak scale. One possible
solution is to consider
the Higgs particle as a composite state of new strongly-interacting
fermions as in technicolor theories. 
Such approaches have however lost their popularity because they tend
to give large positive
contributions to the electro-weak $S$ and $T$ parameters inconsistent
with experimental data coming from
$LEP$ and $SLC$, except for special cases
\cite{Appel}.

Another dynamical symmetry breaking 
scenario has its origin in that the top quark has turned
out to be 
very massive, and in fact quite close to the electro-weak scale. This
could indicate that the Higgs mechanism is closely related to a
top-quark condensate. Models in this direction have provided interesting
insights in the problem of electro-weak symmetry breaking, but they
are usually plagued by various problems. 
Originally they were formulated in terms of four-fermion interactions
of unspecified origin \cite{Man}. In the minimal version they
either do not solve the fine-tuning problem or they predict a top
mass which is much too large \cite{Hill1}. 
In extensions of the minimal
scenario the top mass is also too large, except for the supersymmetric 
or left-right-symmetric extensions. 
This could be an indication that, even though the top quark is an 
important factor in $SU(2)_{L}$ breaking, it is not the only one.
A possible combination of top-mode electro-weak symmetry breaking
and technicolor introduces again the usual problems with the
electro-weak parameters \cite{Hill2}.  

An interesting approach which solves these problems goes in the direction
of introducing new fermions with large electro-weak-invariant masses
\cite{Maekawa}, \cite{Hill3}. These  assist the 
top-quark condensate in breaking $SU(2)_{L}$ and 
simultaneously lead to 
acceptable contributions to the electro-weak parameters due to
the decoupling theorem \cite{Tom}. This paper studies 
a left-right symmetric
model with extra flavor symmetries
which possesses these features, with the additional motivation 
that it can be readily  incorporated into unification schemata
which can in principle produce specific fermion mass hierarchies
and CKM angles.

\section{The model}

It was recently shown in a general context
\cite{Maekawa} that electro-weak-invariant fermion masses
could help in keeping contributions  of new physics to the $S$ and $T$
parameters under control.
These masses
can appear naturally in the theory by introducing, along with new fermions, 
``mirror" fermions with the same quantum numbers but  
opposite handedness. In \cite{Maekawa} these
were introduced in
a technicolor context, but in the present study   
 a left-right and flavor 
symmetric direction is taken. Specifically, mirror families to the 
ordinary Standard Model fermion families are introduced, after extending
their quantum numbers in a left-right symmetric way.
First ideas in this direction appeared quite early \cite{Yang}, but
not in conjunction with gauge-invariant masses. 

In particular, the gauge
group structure $SU(3)_{C} \times SU(2)_{L} \times SU(2)_{R} \times
SU(3)_{F} \times U(1)_{F} \times U(1)_{B-L}$ is considered,  
unbroken at scales on the order of $10^{3}-10^{4}$ TeV.   
The magnitude of these
scales, as will become clear later, is constrained from 
below due to flavor-changing neutral currents (FCNC) and from 
above due to the magnitude of the lightest-family masses. 
The group  $SU(3)_{F}$ unifies the
three Standard-Model families
and the role of the abelian $U(1)_{F}$ group is explained in the following. 
The gauge structure and the new fermions introduced 
have the advantage that, apart from 
restoring the left-right quantum-number 
symmetry missing in the Standard-Model fermions, they can  
be easier embedded in unification schemata, as  will be discussed later. 

Under the above groups,   
the following left-handed fermion representations are introduced:
\begin{equation}
\begin{array}{clcl} 
&{\rm Families} && {\rm Mirror \;\;\; families} \\~ \\
q_{1L}:& ( {\bf 3, \;2, \;1, \;3,} \;\kappa, \;1/3  ) 
& q_{2L}:&( {\bf 3, \;1, \;2, \;\bar{3},} \;-\kappa, \;1/3 ) 
\\~ \\ 
l_{1L}:& ( {\bf 1, \;2, \;1, \;3,} \;\kappa, \;-1  ) 
& l_{2L}:& ( {\bf 1, \;1, \;2, \;\bar{3},} \;-\kappa,\; -1 ) 
\\~ \\ 
q_{1R}^{c}:& ( {\bf \bar{3}, \;1, \;2, \;3,} \;-\kappa, \;-1/3 ) 
& q_{2R}^{c}:& ( {\bf \bar{3}, \;2, \;1, \;\bar{3},} \;\kappa, \;-1/3 )   
\\~ \\
l_{1R}^{c}:& ( {\bf 1, \;1, \;2, \;3,} \;-\kappa, \;1 ) 
& l_{2R}^{c}:& ( {\bf 1, \;2, \;1, \;\bar{3},} \;\kappa, \;1 )   
\end{array}
\end{equation}

\noindent where the subscripts 1,2 indicate whether a fermion is of
Standard-Model type or its mirror, $q$ and $l$ denote
quarks and leptons respectively, $\kappa > 0$ is the
$U(1)_{F}$ charge, and the superscript $c$ denotes
charge conjugation.

The $U(1)_{B-L}$ anomalies are canceled between quarks and 
 leptons and   
the $U(1)_{F}$ anomalies between the fermions and
their mirrors. 
 Moreover, the absence of 
other chiral anomalies in models having 
such a fermion content has been discussed in \cite{Gross}. 
In principle, fermions in such representations could acquire
large gauge-invariant masses on the order of the GUT scale.
However,  the $U(1)_{F}$ coupling 
$\kappa$ is taken to be large enough   
to prohibit the initial formation of 
large $SU(2)_{L} \times SU(2)_{R}$ invariant fermion masses. 
It is worth noting here that the proposed doubling 
 of the fermionic
content in a left-right symmetric context
is typical of models  proposed to provide 
a solution to the strong CP problem \cite{Barr}.

At this
stage, the discrete $L-R$ parity is assumed to be already spontaneously 
broken in such a way that 
the gauge coupling $g_{R}$ corresponding to $SU(2)_{R}$
is larger than the $SU(2)_{L}$ coupling $g_{L}$. 
Such models where $SU(2)_{R}$ and $L-R$ parity break independently
have already been considered in the literature \cite{Moha}.
On the other hand,   
 the family group is assumed to spontaneously break at high energy scales
sequentially down to an abelian group, a process which will
induce  effective four-fermion operators. 
It is then imagined that at a
scale $\Lambda_{R} \approx  500$ GeV the group $SU(2)_{R}$ becomes 
strongly coupled and breaks the abelian gauge group which  
prevented the formation of gauge-invariant masses. 
The fermions which are doublets under $SU(2)_{R}$ acquire therefore 
dynamically gauge-invariant masses. The $SU(2)_{L}$
coupling remains meanwhile weak. 
At lower energies around the electro-weak scale,  
the most attractive of the effective four-fermion operators mentioned above
becomes critical, leading thus to the breaking of the $SU(2)_{R}$
and $SU(2)_{L}$ gauge symmetries. 
One therefore has a scenario  where  
$SU(2)_{R}$ breaks at a low energy scale, 
after it has become strongly coupled.
 The sequence of gauge-symmetry breakings envisaged is graphically 
shown in Fig.1.  

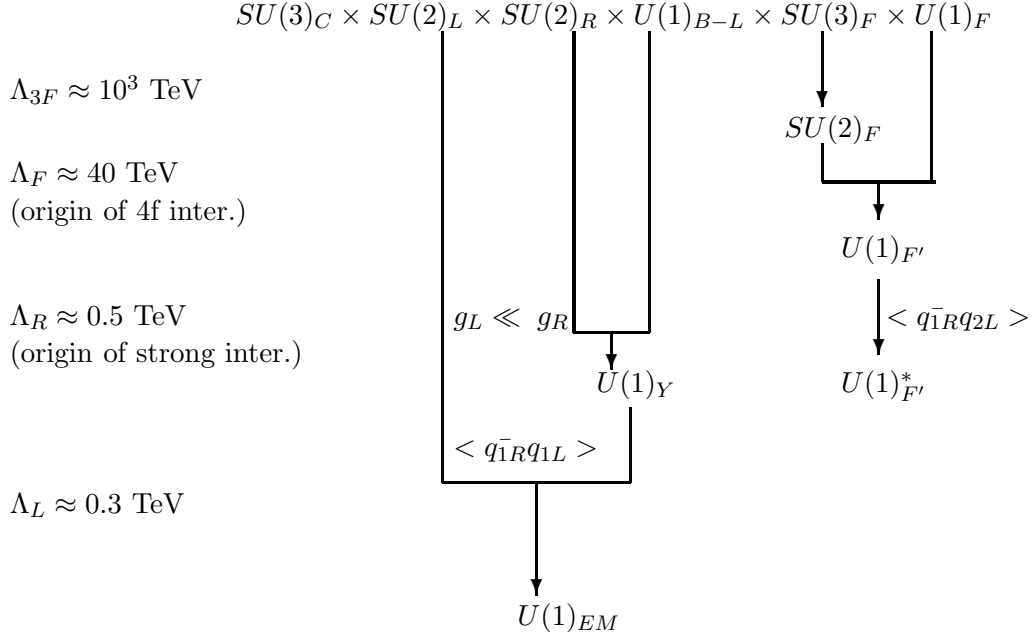
\begin{figure}[t]  
\unitlength0.5cm
\begin{picture}(20,20) \thicklines 
\put(6,20){\makebox(0,0)[bl]{$SU(3)_{C} \times SU(2)_{L} \times SU(2)_{R}
\times U(1)_{B-L} \times SU(3)_{F} \times U(1)_{F}$}} 
\put(21.6,19.8){\vector(0,-1){2}}
\put(0,18){\makebox(0,0)[bl]{$\Lambda_{3F} \approx 10^{3}$ TeV}} 
\put(20.6,17){\makebox(0,0)[bl]{$SU(2)_{F}$}} 
\put(21.6,16.8){\line(0,-1){1}} 
\put(24.5,19.8){\line(0,-1){4}}
\put(21.6,15.8){\line(1,0){3}}
\put(23.1,15.8){\vector(0,-1){1}}
\put(0,15.8){\makebox(0,0)[bl]{$\Lambda_{F} \approx 40$ TeV}} 
\put(0,14.8){\makebox(0,0)[bl]{(origin of 4f inter.)}} 
\put(22.1,13.8){\makebox(0,0)[bl]{$U(1)_{F^{\prime}}$}} 
\put(0,12){\makebox(0,0)[bl]{$\Lambda_{R} \approx 0.5$ TeV}}
\put(0,11){\makebox(0,0)[bl]{(origin of strong inter.)}}  
\put(22.1,10.2){\makebox(0,0)[bl]{$U(1)_{F^{\prime}}^{*}$}}
\put(23.1,13.2){\vector(0,-1){2}}
\put(23.3,12){\makebox(0,0)[bl]{$<\bar{q_{1R}}q_{2L}>$}}  
\put(15,19.8){\line(0,-1){8}}
\put(17,19.8){\line(0,-1){8}}
\put(11.5,19.8){\line(0,-1){12}}
\put(16,11.8){\vector(0,-1){1}}
\put(15,11.8){\line(1,0){2}} 
\put(15.6,10.2){\makebox(0,0)[bl]{$U(1)_{Y}$}}
\put(11.5,7.8){\line(1,0){5}}
\put(16.5,9.8){\line(0,-1){2}}
\put(14,7.8){\vector(0,-1){3}}
\put(11.8,12){\makebox(0,0)[bl]{$g_{L} \ll \; g_{R}$}}
\put(11.8,8.5){\makebox(0,0)[bl]{$<\bar{q_{1R}}q_{1L}>$}}
\put(0,7){\makebox(0,0)[bl]{$\Lambda_{L} \approx 0.3$ TeV}} 
\put(13.5,4){\makebox(0,0)[bl]{$U(1)_{EM}$}}
\end{picture}
\label{fig:diag}
\caption{\small{The sequence of gauge-group breakings of the model
and the energy scales where these take place. 
The $<\bar{q_{1R}}q_{2L}>$ condensate,
originating from the strong $SU(2)_{R}$ interactions at $\Lambda_{R}$,
gives gauge invariant masses to half of the fermions and breaks the 
$U(1)_{F^{\prime}}$  symmetry.
The $<\bar{q_{1R}}q_{1L}> = <\bar{q_{2R}}q_{2L}>$ condensate breaks the
$SU(2)_{L} \times SU(2)_{R}$ and $U(1)_{B-L}$ symmetries. 
It originates from the critical four-fermion interactions coming from the  
$SU(2)_{F}$ gauge group that was broken at $\Lambda_{F}$. 
The superscript $``*"$ indicates
a broken gauge group.}} 
~\\
\end{figure}

A more detailed study of the scenario outlined above is now presented.   
In a first step,  $SU(3)_{F}$ breaks down to 
$SU(2)_{F}$ at a scale $\Lambda_{3F}$, separating
one fermion family from the other two. 
 It will turn out
later that the singlet family under $SU(2)_{F}$ 
is the first and lightest family. 
The scale $\Lambda_{3F}$ should be  
on the order of $10^{3}-10^{4}$ TeV as already explained, 
in order to avoid too large FCNC and   
to get reasonable first generation fermion masses, since  
the massive bosons corresponding to 
the broken generators of $SU(3)_{F}$ are expected 
to feed masses down to first-family 
fermions. 

 After this breaking, 
the  Standard-Model families, together with their
mirror partners, transform with respect to 
 $SU(3)_{C} \times SU(2)_{L} \times SU(2)_{R} \times SU(2)_{F} 
\times U(1)_{F} \times U(1)_{B-L}$ as 
\begin{equation}
\begin{array}{clcl}  
& {\rm Families}& ~~~~~~~~~~~~~~~~~ & 
{\rm Mirror \;\;\; families}  \\
&&& \\
q_{1L}:& (     {\bf 3, \;2, \;1, \;f,} \;\kappa, \;1/3 ) & 
q_{2L}:& ( {\bf 3, \;1, \;2, \;f,} \;-\kappa, \;1/3 ) \\~\\ 
l_{1L}:&     ( {\bf 1, \;2, \;1, \;f,} \;\kappa, \;-1 ) & 
l_{2L}:& ( {\bf 1, \;1, \;2, \;f,} \;-\kappa, \;-1 ) \\~\\  
q_{1R}^{c}:& ( {\bf \bar{3}, \;1, \;2, \;f,} \;-\kappa, \;-1/3 ) & 
q_{2R}^{c}:& ( {\bf \bar{3}, \;2, \;1, \;f,} \;\kappa, \;-1/3 )\\~\\  
l_{1R}^{c}:& ( {\bf 1, \;1, \;2, \;f,} \;-\kappa, \;1 ) & 
l_{2R}^{c}:& ( {\bf 1, \;2, \;1, \;f,} \;\kappa, \;1 ).   
\end{array}
\end{equation}

\noindent where ${\bf f=2}$ for the two heavier 
families and ${\bf f=1}$ for the lightest one.

At a lower scale 
$\Lambda_{F}$,  the  symmetry $SU(2)_{F} \times U(1)_{F}$ 
should 
break spontaneously {\it via} a non-zero vacuum-expectation value having
the right quantum numbers according to the
pattern  $SU(2)_{F} \times U(1)_{F} 
\longrightarrow U(1)_{F^{\prime}}$.
At this point the high-energy physics which generates this breaking 
(for instance by means of a fundamental
Higgs mechanism or of a fermionic composite operator) are left unspecified.
At $\Lambda_{F}$ one has then physics producing 
effective four-fermion 
operators involving second and third family fermions. 
It will turn out later that, 
in order to get the correct electro-weak symmetry
breaking scale, one should have $\Lambda_{F} \approx 40$ TeV. 
What should be kept in mind, however, is that the new physics 
producing this four-fermion term is independent of the
consequences this term implies for lower energy physics and 
that alternative ways to produce it would not affect
the phenomenological results of this work. 

The second and third family quantum numbers 
under $SU(3)_{C} \times SU(2)_{L} \times SU(2)_{R} \times 
U(1)_{F^{\prime }} \times U(1)_{B-L}$  are then given by
\begin{equation}
\begin{array}{clcl}  
& {\rm 2nd \;\; \& \;\;3rd\;\; families}& ~~~~~~~~~~~~~~~~~ & 
{\rm Mirrors \;\;of\;\;2nd\;\;\&\;\; 3rd\;\;families}  \\
&&& \\
q_{1L}:& (     {\bf 3, \;2, \;1,} \;\kappa_{\pm}, \;1/3 ) & 
q_{2L}:& ( {\bf 3, \;1, \;2,} \;-\kappa_{\pm}, \;1/3 ) \\~\\ 
l_{1L}:&     ( {\bf 1, \;2, \;1,} \;\kappa_{\pm}, \;-1 ) & 
l_{2L}:& ( {\bf 1, \;1, \;2,} \;-\kappa_{\pm}, \;-1 ) \\~\\  
q^{c}_{1R}:& ( {\bf \bar{3}, \;1, \;2,} \;-\kappa_{\pm}, \;-1/3 ) & 
q^{c}_{2R}:& ( {\bf \bar{3}, \;2, \;1,} \;\kappa_{\pm}, \;-1/3 ) \\~\\  
l^{c}_{1R}:& ( {\bf 1, \;1, \;2,} \;-\kappa_{\pm}, \;1 ) & 
l^{c}_{2R}:& ( {\bf 1, \;2, \;1,} \;\kappa_{\pm}, \;1 )   
\end{array}
\end{equation}

\noindent where 
$ \kappa_{\pm} = (\kappa \pm 1)/2$ correspond to the 
$U(1)_{F^{\prime}}$ charge $Q_{F^{\prime}} = T_{3F}+Q_{F}/2$
of the second and third families respectively,   
where $T_{3F}$ is an $SU(2)_{F}$ 
generator and $Q_{F}$ is the $U(1)_{F}$ charge. 

As  will be seen in the following, 
the $SU(2)_{F}$ gauge symmetry between the second and
third family plays
a role analogous to the one the
$QCD$-like gauge groups play in top-color models \cite{Hill1}. Its 
breaking induces effective four-fermion operators that will later be 
responsible for the $SU(2)_{R}$ and $SU(2)_{L}$ gauge symmetry
breakings. After Fierz
rearrangement, such a four-fermion term for
the quarks of the second and third generation and their mirrors is 
\begin{eqnarray}
F_{(1,2)}  &= & \frac{\lambda}{\Lambda^{2}_{F}}
({\bar q_{(1,2)R}}q_{(1,2)L})({\bar q_{(1,2)L}}q_{(1,2)R})
\end{eqnarray}

\noindent plus the same term with $L$ and $R$ subscripts interchanged, 
where  $\lambda/\Lambda^{2}_{F}$ is an effective four-fermion coupling.  
The fermion bilinears in both parentheses transform under
$SU(2)_{L} \times SU(2)_{R}$ like  a $({\bf 2,\;2})$.

The next step is connected to the   
assumption made at the beginning, namely that at some high 
energy scale the left-right
parity is broken and that the gauge coupling $g_{R}$ is stronger
than $g_{L}$, where $g_{L,R}$ correspond to $SU(2)_{L,R}$ respectively. 
In fact, it was assumed that at energy scales close to the
$SU(2)_{R}$ characteristic scale $\Lambda_{R}  \approx 500$ GeV, 
the $SU(2)_{R}$ coupling becomes strong enough
to break $U(1)_{F^{\prime}}$ via fermionic condensates.

In order to prevent these condensates from breaking
$QCD$, one has to assume that only two-quark operators like 
$<\bar{q_{1R}}q_{2L}> \approx \Lambda^{3}_{R}/(4\pi)^{2}$ acquire 
non-zero vacuum expectation values with the help of the attractive
$QCD$ interactions and constitute thus the most attractive channel.
Condensates involving leptons correspond to less attractive channels
and are still prohibited by the 
$U(1)_{F^{\prime}}$ gauge symmetry, since for large enough
$U(1)_{F^{\prime}}$ coupling they cannot overcome
the corresponding repulsive interactions \footnote{The $U(1)_{F^{\prime}}$
charge normalization is such that both $\kappa_{+}$ and $\kappa_{-}$
are positive, i.e. $\kappa > 1$. This prohibits also 
the formation of condensates
involving simultaneously second and third generation leptons.}. 
The fermions of the three families that are $SU(2)_{R}$ doublets 
acquire
 $SU(2)_{L} \times SU(2)_{R}$ invariant dynamical masses on the order of 
the $SU(2)_{R}$  scale  $M \approx 
\Lambda_{R} \approx  500$ GeV, 
  while the other half remain 
so far massless. These dynamical masses are equivalent to  
the constituent quark masses in ordinary $QCD$.
The fermion masses get also small contributions from the
effective  four-fermion interactions originating from $SU(2)_{F}$ and
$SU(3)_{F}$. 
 
 Note that these masses are not constrained from 
 above by considerations
concerning Yukawa couplings 
becoming non-perturbative \cite{Csikor}, since their
origin is dynamical and not connected with a symmetry breaking. 
This is novel as regards studies of 
models involving mirror fermions and their phenomenological implications 
\cite{Yang}. 
 It will be interesting to see in the next section how
 the  smallness of the measured $T$ parameter is related to the 
 value of  the dynamical mass $M$. 

The abelian symmetry which protected the fermions from 
acquiring a mass is broken by these condensates. 
Therefore, gauge-invariant mass terms of the form 
${\bar q_{2R}} q_{1L}$ will also appear in the theory. However, they 
will be induced mainly  from the relevant four-fermi operators and will be
on the order of $\Lambda^{3}_{R}/\Lambda^{2}_{F} \approx 0.1$ GeV for 
the second and third generation, and even smaller for the
first generation. 
Interesting mass contributions for the light fermions are thus obtained,
which will be studied elsewhere.
If these light masses are ignored,  
the mass matrix takes the following form 
\begin{eqnarray}
&& \;\;\; q_{2R}^{c} \;\;\; q_{1R}^{c} \nonumber \\ 
\begin{array}{c} q_{1L} \\ q_{2L} \end{array} && 
\left(\begin{array}{cc} 0 & 0 \\ 0 & M \end{array} \right) 
\end{eqnarray}

\noindent for the quarks of the two heavier generations and a similar form
for the leptons and first-generation fermions. The fields in this matrix 
are ordered in a way that will allow
later the direct use of the formalism of \cite{Maekawa}, i.e.
diagonal entries are $SU(2)_{L}$ invariant and
the off-diagonal $SU(2)_{L}$ breaking.  

One should note that the strong
$SU(2)_{R}$ interactions produce mass terms mixing the fermion generations, so
the mass matrices take the above form after diagonalization in 
fermion family space.
This should produce FCNC for the heavy partners of the
Standard-Model fermions, which can in principle be fed down to the known SM 
particles via four-fermion operators. The scales of these 
effective operators are 
 however large enough, in order to avoid problems with FCNC originating from
the broken family groups. Therefore, they are also large enough to 
avoid FCNC in the SM sector coming from the broken $SU(2)_{R}$ group.

For a last step some dynamics are needed 
close to the $SU(2)_{R}$  scale 
$\Lambda_{R}$ which 
leads to the spontaneous breaking    
$SU(2)_{R} \times U(1)_{B-L} \longrightarrow U(1)_{Y}$, where $Y$ is
the usual hypercharge given by  
$Q_{Y}=2 T_{3R} + Q_{B-L}$, with $T_{3R}$ an $SU(2)_{R}$ generator
and $Q_{B-L}$ the $U(1)_{B-L}$ charge. 
This breaking can be achieved 
by a non-zero vacuum-expectation value of either a fundamental
or a composite field  which is a doublet under  
$SU(2)_{R}$ and charged under $U(1)_{B-L}$.  
One of these possibilities will be discussed later, namely how the
breaking of $SU(2)_{R}$  could be due to a fermionic condensate. 
Moreover, since
 $g_{R}$ grows fast at energy scales close to $\Lambda_{R}$, 
it is expected to be much
larger than the $B-L$ coupling $g_{B-L}$ there. Therefore, the hypercharge
gauge coupling $g_{Y}$ at $\Lambda_{R}$ will be approximately equal to 
$g_{B-L}$, since $g_{Y} = 
\frac{g_{R}g_{B-L}}{\sqrt{g^{2}_{R}+g^{2}_{B-L}}}$. 
This relation should constrain the breaking  scale and 
the strength of the coupling  of the  
unifying group from which 
$U(1)_{B-L}$ possibly originates. 

The third family quantum numbers under 
$SU(3)_{C} \times SU(2)_{L} \times U(1)^{*}_{F^{\prime }} 
\times U(1)_{Y}$, 
where the star is a reminder that the gauge group is broken,   
are then 
\begin{equation}
\begin{array}{clcl}  
& {\rm 3rd \;\;family}& ~~~~~~~~~~~~~~~~~ & 
{\rm Mirror \;\;of \;\;3rd \;\; family}  \\
&&& \\
Q_{1L}:& (     {\bf 3, \;2,} \;\kappa_{+}, \;1/3 ) & 
U_{2L}:& ( {\bf 3, \;1, } \;-\kappa_{+}, \;4/3 ) \\~\\ 
& & 
D_{2L}:& ( {\bf 3, \;1, } \;-\kappa_{+}, \;-2/3 ) \\~\\ 
L_{1L}:&     ( {\bf 1, \;2,} \;\kappa_{+},  \;-1 ) & 
N_{2L}:& ( {\bf 1, \;1, } \;-\kappa_{+}, \;0 ) \\~\\  
& & 
E_{2L}:& ( {\bf 1, \;1, } \;-\kappa_{+}, \;-2 ) \\~\\  
U^{c}_{1R}:& ( {\bf \bar{3}, \;1,}  \;-\kappa_{+}, \;-4/3 ) & 
Q^{c}_{2R}:& ( {\bf \bar{3}, \;2,}  \;\kappa_{+}, \;-1/3 ) \\~\\  
D^{c}_{1R}:& ( {\bf \bar{3}, \;1,}  \;-\kappa_{+}, \;2/3 ) & 
& \\~\\ 
N^{c}_{1R}:& ( {\bf 1, \;1,} \;-\kappa_{+}, \;0 ) & 
L^{c}_{2R}:& ( {\bf 1, \;2,}  \;\kappa_{+}, \;1 ) \\~\\  
E^{c}_{1R}:& ( {\bf 1, \;1,}  \;-\kappa_{+}, \;2 ) & 
&    
\end{array}
\end{equation}

\noindent  where 
$Q_{1L,2R} = 
\left(\begin{array}{c} U_{1L,2R}\\ D_{1L,2R} \end{array} \right)$ and
$L_{1L,2R} = 
\left(\begin{array}{c} N_{1L,2R}\\ E_{1L,2R} \end{array} \right)$, 
while the ones for the second family are the same
except for the $U(1)^{*}_{F^{\prime}}$ charges which are $\kappa_{-}$
instead of $\kappa_{+}$.  

Finally, the four-fermion operators $F_{(1,2)}$
involving the 3rd-family up-type
quarks have to be chosen critical, to  form 
$<\bar{U}_{1R} U_{1L}> = <\bar{U}_{2R} U_{2L}> \neq 0$
condensates and break electro-weak symmetry
along the standard 
pattern $SU(2)_{L} \times U(1)_{Y} \longrightarrow U(1)_{EM}$, which  
requires of course a large initial $U(1)_{F^{\prime}}$ coupling
\footnote{This should not pose a problem in principle with a  
Landau pole, since  $U(1)_{F^{\prime}}$ 
is embedded at not too distant energy scales into a non-abelian group.}.   
These fermions therefore acquire $SU(2)_{L}$ breaking masses, which, in
order to reproduce the top quark mass and the weak scale correctly, 
should be on the
order of $m \approx 300$ GeV, as  will be seen in the next section. 
The  gap equations corresponding to the dynamical masses
$m$ and $M$ are diagrammatically shown
in Fig.2. The source of the mass  
$m$ is the effective four-fermion coupling $\lambda/\Lambda_{F}^{2}$, 
which is   
assisted by the $QCD$, hypercharge and $U(1)^{*}_{F^{\prime}}$  couplings 
in a sense of a gauged Nambu - Jona-Lasinio mechanism \cite{gauged}. 
The source of the mass $M$ are, as  has already been seen,
the strong $SU(2)_{R}$ interactions.  

\begin{figure}[t]
\vspace{2cm}
\begin{picture}(6000,6000)

\drawline\fermion[\E \REG](5000,13000)[9000]
\put(\pmidx,\pmidy){\circle*{1000}}
\put(\pbackx,\pbacky){$\; =  \;\frac{\lambda}{\Lambda^{2}_{F}}\;$}
\put(9000,15000){$m$}
\drawline\fermion[\E \REG](19000,13000)[9000]
\put(23500,14000){\circle{2000}}
\put(23500,15000){\circle*{1000}}
\put(23600,15700){$m$}
%
%
\drawline\fermion[\E \REG](5000,5000)[9000]
\put(\pmidx,\pmidy){\circle*{1000}}
\put(\pbackx,\pbacky){$\; = \;g_{R}\;$}
\put(9000,7000){$M$}
\drawline\fermion[\E \REG](19000,5000)[9000]
\put(23500,5000){\circle*{1000}}
\drawloop\gluon[\NE 3](21000,5000)
\put(23600,3500){$M$}
\end{picture}
\caption{\small{The diagrammatic form of the gap equations for 
$m$ and $M$. 
 The wavy line stands for an $SU(2)_{R}$ gauge boson. 
The  corresponding  gauge coupling is denoted by
$g_{R}$ and the four-fermi effective 
 coupling by $\lambda/\Lambda_{F}^{2}$.} }
~\\
\label{fig:gapeq1}
\end{figure}
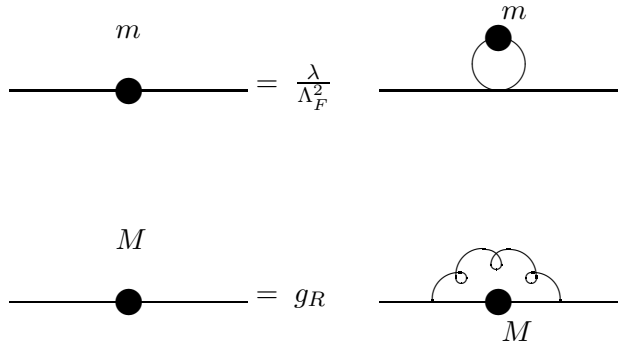

On the other hand, the second family has a smaller charge
under $U(1)^{*}_{F^{\prime}}$ than the third one and 
it is  assumed that its four-fermion interactions are 
not large enough to drive the corresponding gap equations
to criticality.  The same goes for
the down-type quarks and the leptons of the third generation, 
which have smaller hypercharge and no color respectively.  
Lighter family fermions and down type fermions in general 
are expected to acquire their 
masses subsequently by effective operators induced by the broken
$SU(3)_{F}, SU(2)_{F}$ and $SU(2)_{R}$ groups 
respectively. 

The same fermion condensate  
that breaks $SU(2)_{L} \times U(1)_{Y}$ could also be responsible for the  
original $SU(2)_{R} \times U(1)_{B-L}$ 
breaking. However, because the 
characteristic scale of this condensate
is somewhat smaller than the $SU(2)_{R}$ 
scale, non-perturbative contributions push the dynamical masses of the gauge
bosons of $SU(2)_{R}$ up to its characteristic scale
$\Lambda_{R}$. This could in principle
be an economical way of $SU(2)_{R} \times U(1)_{B-L}$ 
breaking avoiding not only too small $SU(2)_{R}$-boson masses
but also
the introduction of additional gauge-symmetry 
breaking mechanisms. In such a scenario
the $U(1)_{Y}$ symmetry would never be essentially 
realised, since it would only be an intermediate technical step
between $U(1)_{B-L}$ and $U(1)_{EM}$. 
It is also worth noting that, 
after inspecting  the $U(1)_{EM}$ quantum numbers of the
mirror families one could qualify them as ``anti-matter". 

The mass matrix for the up-type quarks of the third generation 
and their mirrors, denoted by 
${\cal M_{U}}$,  
takes now the form 
\begin{eqnarray}
&& \;\;\; U_{2R}^{c} \;\;\; U_{1R}^{c} \nonumber \\ 
\begin{array}{c} U_{1L} \\ U_{2L} \end{array} && 
\left(\begin{array}{cc} 0 & m \\ m & M \end{array} \right), 
\label{eq:matrix}
\end{eqnarray} 

\noindent 
while for the mass
matrix of the down-type quarks and their mirrors, denoted by 
${\cal M_{D}}$, one has as before 
\begin{eqnarray}
&& \;\;\; D_{2R}^{c} \;\;\; D_{1R}^{c} \nonumber \\ 
\begin{array}{c} D_{1L} \\ D_{2L} \end{array} && 
\left(\begin{array}{cc} 0 & 0 \\ 0 & M \end{array} \right). 
\end{eqnarray}

After diagonalization therefore, in which the lighter
mass eigenstates are identified with the Standard Model fermions,
a see-saw mechanism \cite{Hill3} 
produces  
small masses for the
SM particles and large masses
for  their partners, in a way that their condensation 
reproduces the weak scale and the top mass correctly.
The large gauge invariant masses of these partners are expected
to damp their
contributions to the electro-weak parameters, as  will be seen in the
following.

\section{Phenomenology} 
For such a model to be phenomenologically viable, it should first of all
be able to
reproduce the known mass hierarchies of the Standard Model fermions and
be consistent with present experimental bounds on new exotic particles.
The new particles introduced should decay fast enough so that cosmological 
problems are avoided. Their decays could however produce interesting 
signals in upcoming experiments like LHC and NLC.
Moreover, the proposed mechanism 
should reproduce the weak scale and not give too large
contributions to FCNC and to 
the S and T parameters. 

The mass $m$ breaks the electro-weak symmetry at a scale $v$.
A rough calculation of the weak scale gives 
\begin{equation}
v^{2} \approx \frac{3}{2\pi^{2}} m^{2} \ln{(\Lambda_{F}/M)} \;, 
\end{equation}

\noindent so
for $\Lambda_{F} \approx $ 40 TeV, $M \approx 500$ GeV and
$m \approx 300$ GeV one gets $v \approx 246$ GeV, as is required.  
The values of $M$ and $m$ are chosen in a way that produces
the correct top quark mass and simultaneously  does not
introduce problems with the $T$ parameter, as  will be seen in the
following.
Note that the factor multiplying the logarithm is twice as 
large as usual, since there are two, independent but equal,
electro-weak-breaking masses $m$. 
It is also worth mentioning
that  such a high scale of four-fermion interactions  
requires a fine-tuning of about $(m/\Lambda_{F})^{2} \approx 10^{-4}$
which is not explained here, but which is typical of similar scenarios
\cite{Hill3}.

The diagonalization of the mass matrix in Eq.\ref{eq:matrix} gives
the eigenvalues $m_{U1} \approx - m^{2}/M$ and $m_{U2} \approx M$
for $m \ll M$. These 
correspond to the eigenvectors
$t$ and $t^{\prime}$, where $t$ is identified with the usual top quark, 
in which case $m_{t} =  - m_{U1}$, while 
$t^{\prime}$ is a heavier partner it mixes with, having 
$m_{t^{\prime}} = m_{U2}$. 
For $m \approx 300$ GeV and $M \approx 500$ GeV 
one gets $m_{t} \approx 175$ GeV. 

\noindent  From the diagonalization one gets 
\begin{equation}
\left(\begin{array}{c}
t_{L,R} \\ t^{\prime}_{L,R}
\end{array}\right) 
= V_{U(L,R)} 
\left(\begin{array}{c}
U_{1L,1R} \\ U_{2L,2R}
\end{array}\right)
\end{equation} 

\noindent while 
\begin{equation}
\left(\begin{array}{c}
b_{L,R} \\ b^{\prime}_{L,R}
\end{array}\right) 
= V_{D(L,R)} 
\left(\begin{array}{c}
D_{1L,1R} \\ D_{2L,2R}
\end{array}\right).  
\end{equation} 

\noindent 
In the above,  $V_{UL} = V_{UR} \approx  
\left(\begin{array}{cc} 
1 & -m/M \\
m/M & 1
\end{array}\right)$, $V_{DL} = V_{DR} = {\bf 1}$, and
\newline ~\\ ${\cal M_{U,D}} = V_{UR,DR}^{\dagger}\left(\begin{array}{cc}
m_{U1,D1} & 0 \\ 
0       & m_{U2,D2} 
\end{array}\right) V_{UL,DR}$, where 
the notation of \cite{Maekawa} is followed closely.
For simplicity the bottom quark mass has been taken equal to zero, so the
corresponding mass matrix is already diagonal with eigenvalues $m_{D1}=0$
and $m_{D2}=M$.   
It is therefore important to note that,
in contrast to the lighter fermion eigenstates, 
the top-quark eigenstate has a non-negligible 
$SU(2)_{L}$-invariant component which could in principle be detectable
in future experiments.

For $S$ and $T$ one obtains then \cite{Maekawa} 
\begin{eqnarray}
S&=& \frac{N}{6\pi}\frac{m^{2}}{M^{2}}
\left(-\frac{4}{3}\ln{(M^{2}/m_{z}^{2})} 
-6\chi(m^{2}/M,M)-M^{2}/m^{2}+2 \right)
\nonumber \\
T&=& \frac{N}{8\pi \sin^{2}{\theta_{w}}m^{2}_{w}}\frac{m^{2}}{M^{2}} 
\left(\theta(M,0)-\theta(M,m^{2}/M)\right), 
\label{eq:Tpar}
\end{eqnarray} 

\noindent where $N$ is the number of contributing new fermion doublets, 
$m_{w,z}$ are the usual $W^{\pm}$ and $Z^{0}$ boson masses and
the functions
$\theta$ and $\chi$ are defined in \cite{Maekawa}.
Note that, in accordance to the decoupling theorem, both $S$ and $T$ tend
to zero as $m/M$ goes to zero.
 A recent fit of experimental data involving $S$ and $T$ gave \cite{Dobre}:
\begin{eqnarray}
S&=&-0.4 \pm 0.55 \nonumber \\
T&=&-0.25 \pm 0.46 \;. 
\end{eqnarray}

\noindent Therefore, one can adjust $M$ in
the present
model so that it gives results consistent with present electro-weak data.

Expanding the $\theta$ and $\chi$ functions in powers of $m/M$, 
 one finally
finds for the $S$ parameter, in leading order and for $M \approx 500$ GeV and 
$N=12$ new $SU(2)_{L}$ doublets: 
\begin{equation}
S = \frac{N}{3\pi}
\frac{m_{t}}{M}\left(\ln{(M^{2}/m^{2}_{t})} 
- 2 \ln{(M^{2}/m_{z}^{2})}/3 -2/3 
\right) \approx -0.44 \;. 
\end{equation}

\noindent 
Note that there
are sizable corrections to this result since $m$ is not much
smaller than $M$, but one should not expect 
qualitative changes of the results when these are included. 

The $T$ parameter is in leading order given by  
\begin{equation}
T = \frac{3 m^{2}_{t}/m^{2}_{w}}{4\pi \sin^{2}{\theta_{w}}}
(2 - \ln{(M^{2}/m^{2}_{t})}) \approx -0.45 \;. 
\end{equation}

\noindent Here it is assumed that $N=3$, i.e. 
only the three (from the three $QCD$ colors) 
``mirror" doublets of the top and bottom quarks contribute.
This is expected, since all other Standard Model quarks have very small masses,
while their ``mirrors" have all masses of order $M$, so their contributions
to the $T$ parameter are vanishingly small, 
as is easily seen from Eq.\ref{eq:Tpar}.  
One can check that, for a given top-quark mass $m_{t}$, 
 values of $M$ too far away from about $500$ GeV
 would yield unacceptable values for the $T$ parameter. 
The relative lightness of the mirror particles that ensues from
this fact provides therefore an accessible
test for the proposed mechanism in experiments like LHC or NLC.
It is important to note at this point that 
the fermion content  used can naturally produce negative values for the
$S$ and $T$ parameters.

We turn now to further phenomenological considerations. 
First of all, it is noted that
the unification scale of the first family with the two heavier ones at 
about $10^{3}-10^{4}$ TeV, as well as the unification scale of 
the second and third families at $\Lambda_{F} \approx 40$ TeV is
too high to 
produce detectable effects like FCNC in present experiments. However, 
the  bosons corresponding to the broken
$U(1)^{*}_{F^{\prime}}$ and $SU(2)_{R}^{*}$ 
could give effects just on 
the border of present experimental constrains on their masses.
Moreover, in this model 
all symmetries are gauged, so there are no light pseudo-goldstone
bosons one should worry about.

Cosmological problems in this scenario are not expected, since 
the mirror families should decay rapidly enough.
Similar to the top quark, mirror quarks
are expected to decay  before
they have the time to hadronize {\it via } four-fermion operators
originating from the broken gauge groups.
They would however be copiously produced at colliders like LHC or NLC.  

The flavor-breaking sequence leads in principle to interesting
hierarchical Yukawa couplings.
Future studies will show 
whether such a model can produce reasonable mass hierarchies, 
CKM matrix
elements and CP violation, as well as reproduce correctly the
measured value of $\sin{\theta_{w}}$. 
In fact, the experimentally known value of $\sin{\theta_{w}}$ 
should allow the prediction of the breaking scale of a possible unification
group from which the assumed group structure  resulted. 
Next, it would be interesting to check if the proposed fermion content 
could lead to some gauge coupling unification at higher energies.
An investigation in this direction \cite{unify} could be updated using the
now known top-quark mass and considering heavier partners for the
Standard-Model fermions, since their masses are gauge invariant, 
while at the same time trying to use a
smaller unifying Pati-Salam breaking scale that can produce reasonably 
large masses for the leptons. 

\section{Discussion}

We have extended the gauge structure and the fermion
content of the Standard Model in a left-right and flavor symmetric way,
by introducing  $SU(2)_{R}$ and $SU(3)_{F}$ gauge groups and mirror fermions
to the ordinary ones. 
A mechanism for electro-weak symmetry breaking was then proposed, which 
reproduces correctly the weak scale and the top-quark mass. 
By giving gauge-invariant dynamical masses to 
half of the fermions due to strong $SU(2)_{R}$ dynamics, 
one is able to naturally produce values for
the $S$ and $T$ parameters in good agreement with experimental
data, a long-standing problem in dynamical symmetry breaking models. 
The masses of the new fermions and bosons are accessible to the
next-generation high-energy experiments, providing therefore a
concrete testing ground for the proposed model. 

The starting point of the model appears even more appealing from a
grand-unification perspective,    
since quark-lepton and family unifying groups give in
principle the possibility to reproduce the observed fermion mass
hierarchies.
In fact, the fermion content used in
this paper fits very nicely in
unification schemata where fermions and vector bosons transform under the
lowest-dimensional representation of an $E_{8}$ group, 
i.e. the adjoint ${\bf 248}$. Other groups could 
also be considered, but the attractiveness of $E_{8}$ comes e.g. from 
string theory \cite{GSW}. If there are no other particles like
fundamental scalars in other representations, $E_{8}$ is supersymmetric
and asymptotically free. 
Under its maximal $SO(16)$ subgroup the particles transform under 
${\bf 128 + 120}$, so if one assumes that $SO(16)$ breaks  
at Planck-scale energies down to its maximal 
$SO(10) \times SU(4)_{F}$ subgroup \cite{Bars},
 where $SU(4)_{F}$ is a fermion family group,   
  the fermions of interest in this paper  transform under the new 
  gauge structure
like $({\bf 16, \;\bar{4}) \; {\rm and} \; (\bar{16}, \;4)}$, i.e.
one has four ordinary and four ``mirror" (or ``conjugate") families. 
The appearance of mirror families is in this context therefore
natural. 
Furthermore, it is intriguing to be able to 
relate the number of fermion families, {\it via} the appearance
of an $SU(4)_{F} \approx SO(6)$ group, with the number of
the (six) compactified dimensions in string theory \cite{GSW}. 

In such a scenario $SO(10)$ then breaks down to 
$SU(4)_{PS} \times SU(2)_{L} \times SU(2)_{R}$, 
where $SU(4)_{PS}$ is a Pati-Salam group \cite{Pati} 
and $SU(4)_{F}$  down to $SU(3)_{F} \times U(1)_{F}$. To avoid
fast proton decay, the $SO(10)$ breaking scale should be larger than
about $\Lambda_{10} \approx 10^{16}$ GeV.
The structure  considered suggests the existence 
of a fourth fermion family and its mirror, 
which is  assumed to acquire a large mass and decouple from the
physics  studied here. An example of how this can be achieved, together with
giving Planck-scale masses also to the other fermions and vector bosons
not needed in this discussion, 
is given in a similar discussion of Ref.\cite{Bars}. 
After the breaking of the Pati-Salam group 
down to $SU(3)_{C} \times U(1)_{B-L}$ 
at around $10^{3}-10^{4}$ TeV,  
 a scale that would allow for reasonable lepton masses to be fed
 down from the $SU(2)_{L}$-breaking up-type quark condensate, 
 one gets the group
structure  and fermion representations 
assumed at the beginning. Other breaking sequences might
also be possible, so this discussion provides only an example of how
one could get elegantly the fermion content  used, and it should not affect
the conclusions  drawn from the proposed mechanism of $SU(2)_{L}$
breaking.

In the past,
a similar fermion content
has been used in connection with a breaking of 
$SO(10)$ down to $SU(5)$, along with a usual Higgs mechanism  \cite{Bagger},
even though in that case the mirror fermions are assumed to 
have electro-weak-breaking masses.  
The motivation for using here
the  $SU(4)_{PS} \times SU(2)_{L} \times SU(2)_{R}$ 
 subgroup of
$SO(10)$ instead of $SU(5)$ 
is that, apart from unifying quarks and leptons in a nice way,
it introduces naturally a  
left-right symmetry which renders the generation of
gauge-invariant masses possible. Moreover, in contrast to \cite{Bagger},
the symmetry that prohibits large gauge-invariant fermion masses 
is here flavor-diagonal, which is due to
the sequential breaking of the family group,  
so there are no problems with FCNC induced by 
the groups $U(1)_{F,F^{\prime}}$. 

Since both $SO(10)$ and $SU(4)_{F}$ are asymptotically free, it is 
conceivable that they self-break {\it via} fermionic 
condensates and tumble down to the assumed gauge structure. 
It would be interesting if the right-handed Standard-Model neutrinos were
involved in such  condensates, because then they would acquire
very large masses and the lightness of their
left-handed partners would be explained by a see-saw mechanism.  
This mechanism would produce neutrino masses small 
enough to provide an MSW solution to the solar neutrino problem
\cite{MSW}. 
A thorough analysis of the attractive channels needed for
such a symmetry
breaking sequence goes however beyond the scope of this paper. 
On the other hand, 
 a Higgs-based mechanism of such a spontaneous breaking sequence, albeit
 in a supersymmetric context, 
is considered for instance in Ref.\cite{Moha}.

It would be nice if the vacuum expectation value that partially
breaks $SO(10)$ 
breaks also the  local $L-R$ discrete symmetry of $SO(10)$, 
explaining thus the large difference of
the couplings $g_{R}$ and $g_{L}$ at low energies. The local 
character of this discrete symmetry and its breaking at scales
higher than the $SU(2)_{R}$ breaking scale avoids also cosmological
problems related to domain-wall formation.
An additional group-theoretic argument 
supporting, but of course not proving,
the simultaneous breaking of $L-R$ parity and $SO(10)$  is the
twofold symmetry of the Dynkin diagram of $SO(10)$, which is a
manifestation of the discrete $L-R$ symmetry and which does not exist 
in the Dynkin diagrams of the subgroups of $SO(10)$  considered here. 

The above arguments all-together show that the specific model 
and the proposed symmetry-breaking sequence  
could very nicely fit into a larger, even more symmetric framework. 
Since there are many possibilities for the dynamics of the theory
at such high energy scales however, one should consider the above just as
pure speculation and only as a hint towards the origin of the gauge structure
and new fermions  introduced,
the representations of which could just as well be taken  {\it ad hoc}. 
In any case, it will be interesting to
see how such a scenario develops as 
its implications are thoroughly investigated in future studies.

\noindent {\bf Acknowledgements} \\
The research of  
G. T. is supported  by an {\it Alexander von Humboldt Fellowship} 
and of M. L. in part by the DFG grant  Li 519/2-2.  
We would like to thank V. Miransky for pointing out
the relevance of Ref. \cite{Maekawa} to this work and J. Manus for
discussions.

\end{document}